\documentclass{jpconf}

\usepackage{amsmath,amssymb}
\usepackage{times}
\usepackage[english]{babel}

\newcommand{\bea}{\begin{eqnarray}}
\newcommand{\eea}{\end{eqnarray}}
\newcommand{\beano}{\begin{eqnarray*}}
\newcommand{\eeano}{\end{eqnarray*}}
\newcommand{\beq}{\begin{equation}}
\newcommand{\eeq}{\end{equation}}

\newcommand{\hs}[1]{\hspace{#1 mm}}

\newcommand{\eps}{\epsilon}

    \def\cH{{\cal H}}

\def\fh{{\mathfrak h}}

\newcommand{\CC}{{\mathbb C}}

\newcommand{\II}{{\mathbb I}}

\newcommand{\wh}[1]{\widehat{#1}}
\newcommand{\wt}[1]{\widetilde{#1}}

\newcommand{\mb}[1]{\hs{4}\mbox{#1}\hs{4}}

\newcommand{\half}{\frac{1}{2}}

\newcommand{\atopn}[2]{\genfrac{}{}{0pt}{}{#1}{#2}}
\newcommand{\up}{\uparrow}
\newcommand{\down}{\downarrow}

\begin{document}
\begin{center}

\title{Generalized Coordinate Bethe Ansatz 
for open spin chains with non-diagonal boundaries}
\author{E Ragoucy\footnote{ragoucy@lapp.in2p3.fr}}
\address{ Laboratoire de Physique Th{\'e}orique LAPTH\\
 CNRS and Universit{\'e} de Savoie.\\
   9 chemin de Bellevue, BP 110, F-74941  Annecy-le-Vieux Cedex}

\begin{abstract}
We introduce a generalization of the original Coordinate Bethe Ansatz that allows to treat the case of
open spin chains with non-diagonal boundary matrices.
We illustrate it on two cases: the XXX and XXZ chains. 

Short review on a joint work with N. Crampe (L2C) and D. Simon (LPMA), see
arXiv:1009.4119, arXiv:1105.4119 and arXiv:1106.3264.
\end{abstract}
\end{center}

\section{Introduction}
The aim of this note is to present a modification of the standard Coordinate Bethe Ansatz so as to deal with open spin chains with non-diagonal boundary matrices. We applied this new method successfully 
 to two models \cite{dam,XXX}, the XXX and XXZ open spin chains, but it clearly should work for other integrable models. To be as simple as possible, we will mainly stick to the XXX model. It will allow us to present in a pedagogical way the generalization we use. The XXZ case is then studied focusing on the differences with the previous case.

The plan of the article is the following. We first introduce in section \ref{sec:XXXclosed} the XXX model with periodic boundary 
conditions. It allows us to present the standard Coordinate Bethe Ansatz and to fix notations. Then, in section \ref{sec:XXXopen} we deal with the open XXX model, with one triangular boundary matrix. This case cannot be dealt with the standard Coordinate Bethe Ansatz, and we use a generalization of it. In section \ref{sec:XXZ} we study the XXZ model with non-diagonal boundary matrices: our method allows to get the eigenfunctions for boundary matrices obeying some constraints, some of them being known, the others new. 
In section \ref{sec:mCBA}, we present  another ansatz we recently introduced, the Matrix Coordinate Bethe Ansatz \cite{NoComCBA}. Together, the generalized Coordinate Bethe Ansatz and this new Ansatz  provide the complete set of eigenvectors and eigenvalues for the XXZ open chain. We conclude in section \ref{sec:conclu}.

\section{XXX model with periodic boundary conditions\label{sec:XXXclosed}}
The XXX spin chain \cite{heis} is one of the most studied integrable models. 
To fix the notations, we start with the XXX model with periodic boundary conditions.
It describes the interaction 
of spins $\half$ on a 1d lattice (of $L$ sites), with Hamiltonian
\begin{eqnarray}
H=\sum_{\ell=1}^{L} h_{\ell,\ell+1}=\sum_{\ell=1}^{L-1} h_{\ell,\ell+1}\ 
{+h_{L,1}}
=\sum_{\ell=1}^{L} \Big(P_{\ell,\ell+1}-\II\otimes\II\Big)\,,
\end{eqnarray}
where $\II$ is the identity and $P_{\ell,\ell+1}$ the 
permutation operator acting on sites $(\ell,\ell+1)$:
\begin{eqnarray}
P=\left(\begin{array}{cccc} 1 & 0 & 0 & 0 \\ 0 & 0 & 1 & 0 \\
0 & 1 & 0 & 0 \\ 0 & 0 & 0 & 1 \end{array}\right)
\qquad P\,( {u}\otimes {v}\,) = {v}\otimes {u}\,.
\label{eq:perm}
\end{eqnarray}
Here and below, we use auxiliary space notation: indices indicate on 
which sites of the chain operators act non trivially. 
For instance 
\begin{eqnarray}
P_{34}=\II\otimes\II\otimes\,P\,\otimes(\II)^{\otimes 
(L-4)}\in\mbox{End}(\cH)\,.
\end{eqnarray}
The interaction is a nearest neighbours interaction, and we set
 $L+1\equiv 1$ (periodic boundary conditions).

The Hamiltonian acts on an 
Hilbert space: $\cH=\big(\CC^2\big)^{\otimes L}$, whose states take the form:
\beq
\underbrace{|\up\rangle\otimes|\up\rangle\otimes|\down\rangle\otimes\cdots
 |\up\rangle\otimes|\down\rangle}_L	
=|\up\up\down\ldots\up\down\rangle
\eeq
 
\subsection{Integrability\label{sec:integ-close}}
It is well-known that the Hamiltonian is integrable and can be obtained from a so-called transfer matrix, see e.g. \cite{fad} and references therein. 
Without going into details, we just remind the steps to get it. One has first to consider the Hamiltonian $\wt H = H +L\,\II$.
It is clear that $\wt H$ and $H$ have  the same eigenfunctions.
Then, one defines the transfer matrix
\begin{eqnarray}
t(\lambda) = tr_0\Big( R_{01}(\lambda)R_{02}(\lambda)\cdots R_{0L}(\lambda) \Big)
\mb{with} R_{k\ell}(\lambda)= \lambda\,\II\otimes\II+P_{k\ell}
\end{eqnarray}
 where $\lambda$ is the spectral parameter, and $P$ is defined in (\ref{eq:perm}). The $R$ matrix $R_{k\ell}(\lambda)$ obeys the celebrated Yang-Baxter equation
 \beq\label{eq:YBE}
R_ {12}(\lambda_1-\lambda_2)\,R_{13}(\lambda_1-\lambda_3)\,R_ {23}(\lambda_2-\lambda_3)\,=\,
R_ {23}(\lambda_2-\lambda_3)\,R_ {13}(\lambda_1-\lambda_3)\,R_ {12}(\lambda_1-\lambda_2)\,.
\eeq
From this relation, it is easy to see that $[t(\lambda)\,,\,t(\lambda')]=0$, $\forall\lambda,\lambda'$, so that upon expansion in $\lambda$, $t(\lambda)$ generates $L$ commuting independent charges. Since $\wt H=\left.\frac{d}{d\lambda}\ln t(\lambda)\right|_{\lambda=0}$, these charges are conserved, which  proves the integrability of the model associated to $\wt H$ and $H$.

\subsection{Coordinate Bethe Ansatz}
We are looking for Hamiltonian eigenfunctions $H\, \Phi = E\, \Phi$. The solution to this problem using the Coordinate Bethe Ansatz has been known for a long time \cite{Gau}.
The starting point is a reference state that is a (zero energy) eigenstate: 
$H\, |\up\ldots\up\rangle = 0$.

Accordingly to this reference state, one can define a general state of the Hilbert space: 
\begin{eqnarray}
&&|x_{1},\dots,x_m\rangle=
|\up\ldots\up\!\!{\raisebox{-1.ex}{$\atopn{\displaystyle\down}{\ x_{1}}$}}\!\!
\up\ldots\up\!\!{\raisebox{-1.ex}{$\atopn{\displaystyle\down}{\ x_{2}}$}}\!\!
\up\ldots\ldots\up\!\!
{\raisebox{-1.ex}{$\atopn{\displaystyle\down}{\ x_{m}}$}}\!\!\up\ldots\up\rangle
\in\big(\CC^2\big)^{\otimes L}
\end{eqnarray}
where $x_{1}<x_2<\dots<x_m$ are the positions of the $m$ spins down of the state.
Note that $m$ is a quantum number: it corresponds to the operator $\frac L2 - S^z$ where 
$S^z=\sum_{\ell=1}^L s^z_\ell$ is the $z$-component of the spin operator (we remind $L$ is the number of sites).

Then, the Coordinate Bethe Ansatz \cite{bethe} is a sort of
plane-waves decomposition with respect to the above basis:
\begin{eqnarray}
 \Phi_m=\sum_{x_{1}<\dots<x_m}\ \sum_{g\in S_m}\
A_g^{(m)}\ e^{i\boldsymbol{k}_g\cdot\boldsymbol{x}}\ 
|x_{1},\dots,x_m\rangle\,.
\end{eqnarray}
$k_j$, $j=1,...,m$ are the plane wave momenta, 
 $S_m$ is the symmetric group ($su(m)$ Weyl group), generated by  
transpositions $\sigma_{j}$, $j=1,\ldots,m-1$ that exchange $k_j$ 
and $k_{j+1}$, and
\begin{eqnarray}
&&\boldsymbol{k}_{g}=(k_{g(1)},\dots,k_{g(m)})\;.
\end{eqnarray}

The coefficients $A_{g}^{(m)}$ are complex numbers to be determined 
such that 
\begin{equation}\label{eq:sch1}
{ H \,\Phi_m = E_m\, \Phi_m\;.}
\end{equation}
We project equation (\ref{eq:sch1}) on the different independent 
vectors ${|x_{1},\dots,x_m\rangle}$ to get constraints on the coefficients 
$A^{(m)}_{g}$. Due to the form of $H$, it is enough to consider three cases only:

\begin{itemize}
\item  all the $x_{j}$'s are far away one from each other 
($1+x_j<x_{j+1}$, $\forall\,j$) and are not on the 
boundary sites 1 and $L$. This case will be called 
{generic}.
\item $x_{j}+1=x_{j+1}$ for one given $j$,
\item $x_{1}=1$ or $x_{m}=L$ (periodicity condition $L+1\equiv 1$).
\end{itemize}
By linearity, more complicated cases just appear as superposition of  
`simple' ones.

These three projections lead to equations that one needs to solve. 
We do not reproduce the calculations here, but just give the solutions.

\paragraph{$\boldsymbol\triangleright$ Calculation of the energy: projection on 
${|x_{1},\dots,x_n\rangle}$ generic}

\begin{eqnarray}
E_m=\sum_{j=1}^m\lambda(e^{ik_j})
\mb{where}
\lambda(u)=u+\frac{1}{u}-2=\frac{(u-1)^2}{u}\,.
\end{eqnarray}

\paragraph{$\boldsymbol\triangleright$ Scattering matrix: projection on 
${|x_{1},\dots,x_{j},x_{j+1}=1+x_{j},\dots,x_m\rangle}$}
\ \\
It provides the scattering 
matrix between pseudo-excitations. 
\begin{eqnarray}
A^{(m)}_{g\sigma_j}
&=&S\!\left(e^{ik_{g(j)}},e^{ik_{g(j+1)}}\right)\,A^{(m)}_{g}\,,
\label{eq:Aclosed}
\\
{S(u,v)}&=&-\frac{2v-uv-1}{2u-uv-1}
\end{eqnarray}
Eq. (\ref{eq:Aclosed}) allows to express all the coefficients $A^{(m)}_{g}$ in term of a single one, say 
$A^{(m)}_{id}$, where $id$ is the identity in $S_m$.

\paragraph{$\boldsymbol\triangleright$ Bethe equations: projection on 
${|x_{1}\dots,x_{m-1},L\rangle}$}
\ \\
This last constraint consists in the quantization of the pseudo-excitation
momenta $k_{j}$ since the 
system is in a finite volume. 
\begin{eqnarray}
&& \prod_{\substack{\ell=1 \\ \ell\neq j}}^m 
S(e^{ik_\ell},e^{ik_j})=e^{iLk_j}\,
 \mb{for} 1\leq j \leq m
\end{eqnarray}

With these three projections, one gets all the relevant physical information for the model, and obtains the eigenfunctions of the Hamiltonian. 
 
\section{XXX model with boundaries\label{sec:XXXopen}}
The open XXX model has Hamiltonian 
\begin{eqnarray}
H=B_{1}^+ +H_{bulk}+ B_{L}^- \mb{where} 
H_{bulk}=\sum_{\ell=1}^{{L-1}} h_{\ell,\ell+1}
=\sum_{\ell=1}^{{L-1}} \Big(P_{\ell,\ell+1}-\II_{4}\Big)
\end{eqnarray}
with boundary matrices $B^\pm$ that we choose of the form
\begin{eqnarray}
B^+=\left(\begin{array}{cc} \alpha & {\mu} \\ 0 & \beta 
\end{array}\right)
\mb{and} B^-=\left(\begin{array}{cc} \gamma & 0 \\ 0 & \delta 
\end{array}\right)\,.
\end{eqnarray}

This Hamiltonian describes the interaction of spins 
(up or down) among themselves, and with two boundaries described by 
the matrices $B^\pm$. These matrices preserve integrability of the model (see below).

Let us stress that the boundary matrices corresponds to a new case, the $B^+$ matrix being triangular (when $\mu\neq0$). 
This means that 
the left boundary can now flip the spin $\down$ to $\up$. This has drastic consequences as we shall see.

\subsection{Gauge transformations}
Obviously, any Hamiltonian $H'$ related to $H$ by a gauge transformation, 
\beq
H'=\underbrace{U\otimes U\otimes\cdots\otimes U}_L\,H\,\underbrace{U^{-1}\otimes U^{-1}\otimes\cdots\otimes U^{-1}}_L
\eeq
 will have the same spectrum as $H$, and their  eigenfunctions will be related in an obvious way. 

Since $H_{bulk}$ is invariant under these gauge transformations,
our approach is valid for any boundary matrices deduced from $B^\pm$ by a gauge transformation 
$K^\pm=U\,B^\pm\,U^{-1}$. 

In particular, for the case
\beq
U= i\,\left(\begin{array}{cc} 0 & 1 \\ 1 & \nu\end{array}\right)
\eeq
one gets lower triangular matrices
\begin{eqnarray}
K^+=\left(\begin{array}{cc} \beta & 0 \\ \mu+\nu(\alpha-\beta) & \alpha 
\end{array}\right)
\mb{and} K^-=\left(\begin{array}{cc} \delta & 0 \\ \nu(\gamma-\delta) & \gamma 
\end{array}\right)\,.
\end{eqnarray}

On the contrary, when $\alpha=\beta$ and $\delta\neq\gamma$, it is not possible to find a gauge transformation 
that diagonalizes $B^+$ while keeping $B^-$ diagonal, since the two matrices do not commute.

It should be also clear that the same treatment can be done when $B^+$ is diagonal and $B^-$ is triangular.

\subsection{Integrability and connection with reflection algebra}
For those familiar with the so-called reflection equation\cite{cherednik,sklyanin}, let us note that the matrices $B^\pm$ do not obey this equation. However, they are connected to such 'reflection matrices' in the following way. One first has to make them traceless, using the identity matrix (which does not change the form of the eigenfunctions) and consider the Hamiltonian
\begin{eqnarray}
\wt H = H - \half (trB^-)\,\II_1 - \half (trB^+)\,\II_L+(L-1)\,\II\equiv H+\big(L -1-\half (trB^-)- \half(trB^+)\big)\,\II
\end{eqnarray}
where the indices 1 and $L$  (that are in fact not relevant when considering the identity matrix) are explicited for obvious reasons.
Then, the two new matrices $\wt B^\pm=B^\pm - \half(tr B^\pm)\,\II$ are connected to 'reflection matrices' through
\bea
K^\pm(\lambda)=\II+\lambda\,\wt B^\pm
\end{eqnarray}
where $\lambda$ is the spectral parameter. These 'reflection matrices' obey
\begin{eqnarray}\label{eq:BYBE}
R_{12}(\lambda_1-\lambda_2)\,K^\pm_1(\lambda_1)\,
R_{12}(\lambda_1+\lambda_2)\,K^\pm_2(\lambda_2)
=
K^\pm_2(\lambda_2)\,R_{12}(\lambda_1+\lambda_2)\,
R_{12}(\lambda_1-\lambda_2)\,K^\pm_1(\lambda_1)
\end{eqnarray}
where $R_{12}(\lambda)$ has been defined in section \ref{sec:integ-close}.
Again, from the Yang-Baxter equation (\ref{eq:YBE}) and the reflection equation (\ref{eq:BYBE}), one proves that the transfer matrix \cite{sklyanin}
\begin{eqnarray}
t(\lambda) = tr_0\Big( K^+(\lambda)\,R_{01}(\lambda)\cdots R_{0L}(\lambda)\,K^-(\lambda)
\,R_{0L}(-\lambda)\cdots R_{01}(-\lambda) \Big)
\end{eqnarray}
obeys $[t(\lambda)\,,\,t(\lambda')]=0$, $\forall\lambda,\lambda'$. Since 
one has $\wt H=\left.\frac{d}{d\lambda}t(\lambda)\right|_{\lambda=0}$, this proves the integrability of the model associated to $\wt H$.

\subsection{Generalized Coordinate Bethe Ansatz}
Again, as for the periodic case, the starting point is a 
reference state:
$H\, |\up\ldots\up\rangle = (\alpha+\gamma)\, |\up\ldots\up\rangle$.
Note that even when $\mu\neq0$ this state is a reference state, while $ |\down\ldots\down\rangle$ is not anymore.

$\boldsymbol\triangleright$ When $\mu=0$ (diagonal boundaries) the boundaries do not 
modify the spin (no flip) and one can use the "usual" Coordinate Bethe Ansatz:
\begin{eqnarray}
\Phi_m=\sum_{x_{1}<\dots<x_m}\ \sum_{{g\in BC_m}}\
A_g^{(m)}\ e^{i\boldsymbol{k}_g\cdot\boldsymbol{x}}\ 
|x_{1},\dots,x_m\rangle\,,
\end{eqnarray}
$BC_m$ is the $B_{m}$ Weyl group, generated by  
the symmetric group $S_m$ and the reflection $R_{1}$ exchanging $k_{1}$ and 
$-k_{1}$.

$\boldsymbol\triangleright$ When $\mu\neq0$, one has to modify the Ansatz
\begin{eqnarray}
\Psi_n= {\sum_{m=0}^n}\ {\sum_{x_{m+1}<\dots<x_n}\ \sum_{{g\in G_m}}\
A_g^{(n,m)}\ e^{i\boldsymbol{k}^{(m)}_g\cdot\boldsymbol{x}^{(m)}}\ 
|x_{m+1},\dots,x_n\rangle}\,,
\end{eqnarray}
where $G_m=BC_n/BC_m$ and $\boldsymbol{k}^{(m)}_g\cdot\boldsymbol{x}^{(m)}=
\sum_{j=m+1}^n k_{g(j)} x_j$.

Let us stress that in this model the number of pseudo-excitations $\down$ is \underline{not} conserved, although the model is still integrable.

The coefficients $A_{g}^{(n,m)}$ are all determined (but one) by   
\begin{eqnarray}
 H \,\Psi_n = E_n\, \Psi_n\;.
\end{eqnarray}

We project this equation on states $|\vec{x}\,\rangle$ with:
\begin{itemize}
\item  $(x_{1},x_2,...,x_n)$  generic ($1+x_j<x_{j+1}$, $\forall\,j$) 
\item $x_{j}+1=x_{j+1}$ for some $j$
\item $x_{n}=L$
\item $x_{1}=1$
\item $(x_{m+1},...,x_n)$ generic ($m>0$)
\end{itemize}
As in the periodic case, these projections lead to equations that have to be solved. We do not reproduce the (rather lengthy) calculations, and only give the results. We refer to \cite{XXX} for details on the calculation.
The results contain the relevant physical information of the model.

\paragraph{$\boldsymbol\triangleright$ Calculation of the energy: projection on 
${|x_{1},\dots,x_n\rangle}$ generic}

\begin{eqnarray}
E_n=\alpha+\gamma+\sum_{j=1}^n\lambda(e^{ik_j})
\mb{where}
\lambda(u)=u+\frac{1}{u}-2=\frac{(u-1)^2}{u}\,.
\label{eq:Eopen}
\end{eqnarray}
Note that it has a "bulk part" similar to the periodic case, and a boundary contribution that is independent of $\mu$.

\paragraph{$\boldsymbol\triangleright$ Scattering matrix: projection on 
${|x_{1},\dots,x_{j},x_{j+1}=1+x_{j},\dots,x_n\rangle}$}

\begin{eqnarray} 
A^{(n,0)}_{g\sigma_j} &=&
S\!\left(e^{ik_{g(j)}},e^{ik_{g(j+1)}}\right)\,A^{(n,0)}_{g}
\mb{where}
 S(u,v) \,=\, -\frac{2v-uv-1}{2u-uv-1}
\end{eqnarray}
It is similar to the periodic case since the 
boundaries are not involved in this process.

\paragraph{$\boldsymbol\triangleright$ Reflection coefficient for the left boundary: projection on 
${|1,x_{m+1}\dots,x_n\rangle}$}

\begin{eqnarray}
&&A^{(n,0)}_{gR_{1}} = R(e^{ik_{g1}})\ A^{(n,0)}_{g}
\mb{where}
R(z)\,=\,-z^2\,\frac{1-\frac1z+\beta-\alpha}{1-z+\beta-\alpha}
=\frac{r_+(1/z)}{r_+(z)}\,,
\\
&&r_+(z) =
-\frac{(z-1)(1-z+\beta-\alpha)}{z(1+z)}\,.
\end{eqnarray}
This equation is specific to open case (i.e. in presence of a boundary), but is valid whatever the boundary matrices are (diagonal or not).

\paragraph{$\boldsymbol\triangleright$ Transmission coefficient: projection on 
${|x_{m+1}\dots,x_n\rangle}$}

\begin{eqnarray}
&&A^{(n,m)}_{g} = T^{(m)}(e^{ik_{g(1)}},...,e^{ik_{g(m)}})\ A^{(n,m-1)}_{g}
\\
&& T^{(m)}(z_1,...,z_m)
=\frac{{\mu}}{r_+(z_m)\,\prod_{j=1}^{m-1}a(z_{m},z_j)\,a(z_{j},1/z_m)}
\\
&& a(z_1,z_2)=i\,\frac{2z_2-z_1 z_2-1}{z_1 z_2-1}\,.
\end{eqnarray}
This equation is specific to the case of triangular boundary matrices. It is new with respect to the case of diagonal boundary matrices.
It relates the coefficients $A^{(n,m)}_{g}$ with different $m$'s: it shows that the number of spins down cannot be conserved. In other words we have a system where the number of "pseudo-particles" (spin down) is not conserved, while the model is still integrable.

\paragraph{$\boldsymbol\triangleright$ Bethe equations: projection on 
${|x_{1}\dots,x_{n-1},L\rangle}$} 

\begin{eqnarray}
&& \prod_{\substack{\ell=1 \\ \ell\neq j}}^n 
S(e^{ik_\ell},e^{ik_j})\,S(e^{-ik_j},e^{ik_\ell})
=e^{2iLk_j}\,
\frac{r_+(e^{ik_{j}})\,r_-(e^{ik_{j}})}
{r_+(e^{-ik_{j}})\,r_-(e^{-ik_{j}})} \quad\qquad 1\leq j \leq n
\quad\\
&& r_{-}(z) = \frac{z-1}{z+1}\,(1-z+\delta-\gamma)\,.
\end{eqnarray}

Note that, as in eq. (\ref{eq:Eopen}), the Bethe equations do not depend on $\mu$. 
This proves that the eigenvalues are the same as the ones of the model associated to diagonal boundary matrices. This correspondence ensures that the eigenvalues are real although the Hamiltonian is not Hermitian. 

Let us stress that although the eigenvalues do not depend on $\mu$, the eigenvectors do. Hence the physical properties of the model are different.

\section{Generalization to XXZ model with boundaries\label{sec:XXZ}}
The resolution of XXZ model with non-diagonal matrices shares the same ideas but with extra new features. Thus, we will not describe the approach in details and rather focus on these extra features, referring to \cite{dam} for details. However, to stick to the presentation done for XXX model, we will use XXZ notation, instead of the ASEP one used in \cite{dam}. The explicit form of the transformation 
relating the two notations can be found in e.g. \cite{dGE}. 

The Hamiltonian we consider has the form
\bea 
H &=& \wh B_1 + B_L -\half\sum_{j=1}^{L-1}\Big\{\sigma^x_j\sigma^x_{j+1}
+\sigma^y_j\sigma^y_{j+1}+\Delta(\sigma^z_j\sigma^z_{j+1}-\II)-\fh\,(\sigma^z_j-\sigma^z_{j+1})\Big\}\,,
\label{eq:hamxxz}\\
\Delta &=& \frac12(Q+Q^{-1}) \mb{and} \fh\,=\, \frac12(Q-Q^{-1})\\
\widehat B&=&
\left(\begin{array}{c c}
\alpha  & -\mu\gamma e^{-s}  
\\
 -\frac{\alpha}{\mu}e^{s}  & \gamma
\end{array}\right)
\quad\mbox{ and }\quad
B\,=\,
\left(\begin{array}{c c}
\delta  & -\mu\beta Q^{L-1}  
\\
-\frac{\delta}{\mu Q^{}L-1} & \beta
\end{array}\right)
\label{eq:Bxxz}
\eea
where $\sigma$ are the usual Pauli matrices and $\mu$ is a free parameter.

\subsection{Constraints between the (non-diagonal) boundary matrices}
\ \\
To solve the XXX model, we considered one triangular and one diagonal boundary matrix (up to gauge transformations). In the same way, when one deals with the XXZ model, we need to consider special (non-diagonal) matrices of the form (\ref{eq:Bxxz}). However, in addition to this special form, they have to obey some constraint relations:
\bea
&&\prod_{\eps,\eps'=\pm}\Big(c_\eps(\alpha,\gamma)\,c_{\eps'}(\beta,\delta)-Q^{L-1-n}\,e^{-s}\Big)=0
\label{eq:c1-1}
\mb{with} c_+(u,v)=\frac uv \quad \mbox{and}\quad c_-(u,v)=1\qquad
\eea
or
\bea
\prod_{\eps,\eps'=\pm}\Big(\wt c_\eps(\alpha,\gamma)\,\wt c_{\eps'}(\beta,\delta)-Q^{-n}\,e^{s}\Big)&=&0
\label{eq:c2-1}\\
\mb{with} \wt c_\pm(u,v)&=&\frac{Q^{-1}-Q+v-u\pm\sqrt{(Q^{-1}-Q+v-u)^2+4uv}}{2u}\quad
\label{eq:c2-2}
\eea
where the integer $n$ corresponds to the eigenfunction $\Psi_n$ on which they act.
 Indeed, the first choice of constraints (\ref{eq:c1-1}) has to be related to the
 original approach \cite{nepo} that allowed to compute eigenvalues for XXZ model with non-diagonal boundary matrices using fusion relations. 
 The second choice (\ref{eq:c2-1})-(\ref{eq:c2-2}) corresponds to new constraints. In both cases, we computed the eigenvalues and the eigenfunctions of the corresponding model. Below we present some results for the first choice of constraints, the complete treatment being done in \cite{dam}.
 
\subsection{Basis vectors depend on which site they are} 
\ \\
The usual spin up and spin down vectors used in the XXX models have now to be replaced by the following vectors
\begin{eqnarray}
|\up\rangle_\ell \ \to\ |u_k\rangle_\ell= \begin{pmatrix} 1 \\  \mu\,Q^{\ell-1}\,u_k \end{pmatrix}_\ell
\quad;\quad
|\down\rangle_\ell \ \to\ |d_k\rangle_\ell= \begin{pmatrix} 1 \\  \mu\,Q^{\ell-1}\,d_k\end{pmatrix}_\ell
\qquad \ell=1,...,L
\label{eq:vectors}
\end{eqnarray}
Remark that they depend on the site $\ell$ where they stand and also of extra parameters $u_k,d_k$. This site dependence has to be related to the local gauge transformations \cite{cao} that are used to construct the Algebraic Bethe Ansatz for XXZ model with non-diagonal matrices.

Then, a generic Hilbert space vector is defined by
\begin{eqnarray}
|x_{m+1},\ldots,x_n\rangle=
|u_{m+1}..u_{m+1} {\raisebox{-0.81ex}{$\atopn{\displaystyle d_{m+1}}{x_{m+1}}$}} 
u_{m+2}\ldots u_{n}
{\raisebox{-.81ex}{$\atopn{\displaystyle d_{n}}{x_{n}}$}}u_{n+1}.. u_{n+1}\rangle
\label{eq:genVxxz}
\end{eqnarray}
and the parameters $u_{m+1},...,u_n$ and $d_{m+1},...,d_n$ are fixed by the generalized Coordinate Bethe Ansatz. In particular they obey the relations $u_{\ell+1}=Q^{-1}u_\ell$ and $d_{\ell+1}=Q^{-1}d_\ell$.
 
\subsection{Telescoping terms appear}
\ \\
When local Hamiltonians $h_{\ell,\ell+1}$ act on generic vectors (\ref{eq:genVxxz}), they make appear new vectors
$|t\rangle = \left(\begin{array}{c} 1\\0\end{array}\right)$ that are not of the form (\ref{eq:vectors}):
 \begin{eqnarray}
h_{12} |d\rangle\otimes |u\rangle &=&  |d\rangle\otimes |u\rangle - Q  |u\rangle\otimes |d\rangle 
+ (Q-Q^{-1}) |d\rangle\otimes |t\rangle
\\
h_{12} |u\rangle\otimes |d\rangle &=&   |u\rangle\otimes |d\rangle - Q^{-1}  |d\rangle\otimes |u\rangle
- (Q-Q^{-1}) |d\rangle\otimes |t\rangle
\\
h_{12} |d\rangle\otimes |d\rangle &=&   (Q - Q^{-1})  \Big\{|d\rangle\otimes |t\rangle
-  |t\rangle\otimes |d\rangle \Big\}
\\
h_{12} |u\rangle\otimes |u\rangle &=&   0  
\end{eqnarray}
In view of these relations, one could be tempted to think that the basis (\ref{eq:genVxxz}) is not suited for the study of the XXZ Hamiltonian.
However, this is only true for local Hamiltonians $h_{\ell,\ell+1}$. On the general Hamiltonian (\ref{eq:hamxxz}), these new vectors appear with alternating signs, so that their only contribution to the total Hamiltonian is on the first and last site, where they are used to diagonalize the boundary matrices. 
\\ \ \\
Apart from these three modifications, the generalized Coordinate Bethe Ansatz for the XXZ model follows the same steps as for the XXX one.

\section{Completeness and Matrix Coordinate Bethe Ansatz\label{sec:mCBA}}

In the case of XXX model, it has been shown that the Coordinate Bethe Ansatz provides the complete set of eigenvectors for periodic boundary conditions
\cite{kiri}. It is also believed to be complete for open diagonal boundary conditions. For triangular boundary matrix, the spectrum being the same as for diagonal ones,  the set should be complete too \cite{XXX}.

For XXZ model, it is known that the generalized Coordinate Bethe Ansatz do not provide all the eigenvectors. 
For instance, by numerical investigations \cite{neporav}, it has been established that the whole spectrum is given by two different types of Bethe equations.
The present method provides the eigenvalues and  
eigenvectors corresponding to only one type of Bethe equations. 
Remark that since the Hamiltonian is not Hermitian, the right and left eigenvectors are different. Thus,
using the Coordinate Bethe Ansatz on left vectors leads to another set of eigenvalues. Together,  the "right" and "left" eigenvalues  generate the complete spectrum. However, in this way, one constructs only "half" of the eigenvectors for each "side" (right or left vectors). Although this generalized Coordinate Bethe Ansatz is not enough to obtain all the vectors, it has
the advantage of  giving an interpretation of the number $n$ 
entering in the constraint: it is the maximal number of pseudo-excitations in the Ansatz. \ \ \\
To get a complete set of eigenvectors, one needs to use another ansatz: in few cases, one can use the so-called
Matrix ansatz \cite{DEHP} (used in Statistical Physics), but in general it is not sufficient since it provides only one eigenvector. 

In \cite{NoComCBA}, we developed a Matrix Coordinate Bethe Ansatz, that is a
mixing of generalized coordinate Bethe ansatz and of Matrix ansatz. More precisely, it is a 
non-commutative generalized coordinate Bethe ansatz, where the entries in (\ref{eq:genVxxz}) now belong to an algebra 
(very closed to the one introduced in \cite{DEHP}) acting in an additional auxiliary space. 
In this framework, the Matrix ansatz eigenvector appears as a new vacuum on which we build this 
non-commutative generalized coordinate Bethe ansatz.
Numerical studies indicate the spectrum is then complete.
For more details, we refer to the recent work \cite{NoComCBA}.

\newpage

\section{Conclusion\label{sec:conclu}}
We have shown a generalization of the Coordinate Bethe Ansatz that allows to take into account the case of non-diagonal boundary matrices. The Ansatz has been applied to XXX and XXZ open spin chains, but it should also work on different integrable models. In the case of XXZ model, the Ansatz allows to recover and generalize the constraints found with different methods. However, the case of fully general boundary matrices remains to be done.

In the case presented here, completeness of the Ansatz is ensured by the introduction of another Ansatz, the Matrix Coordinate Bethe Ansatz. A synthetical presentation of both Ans\"atzen is also lacking for the moment.

Finally, let us stress that these Ans\"atzen can also be applied to open spin chains built on algebras of higher rank. 

Works are in progress on these subjects.

\end{document}